\documentclass[draftcls,12pt,onecolumn,dvipdfm]{IEEEtran}
\usepackage{mathrsfs}
\usepackage{amsmath}
\usepackage{amssymb}
\usepackage{enumerate}
\usepackage{color,xcolor}
\usepackage{graphicx}
\usepackage{subfigure}
\usepackage{pifont}
\usepackage{algorithmic}

\ifCLASSINFOpdf

\else

\fi

\newtheorem{theorem}{Theorem}
\newtheorem{proposition}{Proposition}
\newtheorem{corollary}{Corollary}
\newtheorem{definition}{Definition}
\newtheorem{lemma}{Lemma}
\newtheorem{algorithm}{\textbf{Algorithm}}

\usepackage[square,numbers,sort&compress]{natbib}

\usepackage{ifpdf}

\begin{document}

\title{Upper Bounds On the ML Decoding Error Probability of General Codes over AWGN Channels}

\author{Qiutao~Zhuang,
        Jia~Liu,
        and~Xiao~Ma,~\IEEEmembership{Member,~IEEE}
\thanks{The authors are with the Department
of Electronics and Communication Engineering, Sun Yat-sen
University, Guangzhou 510006, China. J. Liu is also with the College
of Information Science and Technology, Zhongkai University of
Agriculture and Engineering, Guangzhou 510225, China. (Email:
maxiao@mail.sysu.edu.cn)}}


\maketitle

\begin{abstract}
In this paper, parameterized Gallager's first bounding
technique~(GFBT) is presented by introducing nested Gallager
regions, to derive upper bounds on the ML decoding error probability
of general codes over AWGN channels. The three well-known bounds,
namely, the sphere bound~(SB) of Herzberg and Poltyrev, the
tangential bound~(TB) of Berlekamp, and the tangential-sphere
bound~(TSB) of Poltyrev, are generalized to general codes without
the properties of geometrical uniformity and equal energy. When
applied to the binary linear codes, the three generalized bounds are
reduced to the conventional ones. The new derivation also reveals
that the SB of Herzberg and Poltyrev is equivalent to the SB of
Kasami~{\em et~al.}, which was rarely cited in the literatures.
\end{abstract}


%
\IEEEpeerreviewmaketitle

\section{Introduction}\label{introduction}
%
%
%
%

\IEEEPARstart{I}{n} most scenarios, there do not exist easy ways to
compute the exact decoding error probabilities for specific codes
and ensembles. Therefore, deriving tight analytical bounds is an
important research subject in the field of coding theory and
practice. Since the early 1990s, spurred by the successes of the
near-capacity-achieving codes, renewed attentions have been paid to
the performance analysis of the maximum-likelihood (ML) decoding
algorithm. Though the ML decoding algorithm is prohibitively complex
for most practical codes, tight bounds can be used to predict their
performance without resorting to computer simulations. As mentioned
in~\cite{Sason06}, most bounding techniques have connections to
either the 1965 Gallager
bound~\cite{Duman98,Duman98a,Shulman99,Twitto07} or the 1961
Gallager
bound~\cite{Berlekamp80,Kasami92,Kasami93,Sphere94,TSB94,Divsalar99,Divsalar03,Yousefi04,Yousefi04a,Mehrabian06,Ma10,Ma11,Ma13}
based on Gallager's first bounding technique~(GFBT). However, most
previously reported upper bounds are focusing on binary linear
codes.

For binary linear codes modulated by binary phase shift keying
(BPSK), there are two main properties, which are {\em geometrical
uniformity} and {\em equal energy}. The geometrical uniformity
allows us to make an assumption that the all-zero codeword is the
transmitted one, while the property of equal energy is critical to
derive the tangential bound~(TB)~\cite{Berlekamp80} and the
tangential-sphere bound~(TSB)~\cite{TSB94}. For general codes
without these two properties, performance analysis becomes more
difficult than that for binary linear codes.

In this paper, we present parameterized GFBT by introducing nested
Gallager regions to derive upper bounds on the ML decoding error
probability of general codes over AWGN channels. The main
contributions as well as the structure of this paper are summarized
as follows.

\begin{enumerate}
\item We present in Sec.~\ref{sec2} the parameterized GFBT for general codes. We also present a necessary and sufficient condition
on the optimal parameter, and a sufficient condition~(with a simple
geometrical explanation) under which the optimal parameter does not
depend on the signal-to-noise ratio~(SNR).
\item  Within the general framework based on the introduced nested Gallager
regions, three existing upper bounds, the sphere bound~(SB) of
Herzberg and Poltyrev~\cite{Sphere94}, the tangential bound~(TB) of
Berlekamp~\cite{Berlekamp80} and the tangential-sphere bound~(TSB)
of Poltyrev~\cite{TSB94}, are generalized in Sec.~\ref{sec4} to
general codes without the properties of geometrical uniformity and
equal energy. The three upper bounds are then applied to binary
linear codes and reduced to the conventional ones. The new
derivation also reveals that the SB of Herzberg and Poltyrev is
equivalent to the SB of Kasami {\em
et~al.}~\cite{Kasami92}~\cite{Kasami93}, which was rarely cited in
the literatures.
\item We use in Sec.~\ref{sec5} terminated trellis codes~\cite{Caire98a} to illustrate how to
calculate the parameterized Gallager first bounds on the frame-error
probability. Numerical results are also presented in
Sec.~\ref{sec5}. Sec.~\ref{conclusion} concludes this paper.
\end{enumerate}

\section{The Parameterized Gallager's First Bounds}\label{sec2}

\subsection{General Codes}
A {\em general code} $\mathcal{C}(n,M)\subset \mathbb{R}^{n}$, in
this paper, means a set that contains $M$ $n$-dimensional real
vectors (referred to as codewords). The squared Euclidean distance
between a codeword $\underline s$ and the origin point $\mathbb{O}$
of the $n$-dimensional space, denoted by $\|\underline s\|^2$, is
also referred to as the energy of this codeword. If all codewords
have the same energy, we say that the code has the property of equal
energy.

Given a codeword $\underline{s}$, we denote
$A_{\delta_d|\underline{s}}$ the number of codewords having the
Euclidean distance $\delta_d$ with $\underline{s}$. We define
\begin{equation}\label{Ad_Ads}
A_{\delta_d} \stackrel{\Delta}{=} \sum_{\underline{s}}{\rm
Pr}\{\underline{s}\}A_{\delta_d|\underline{s}},
\end{equation}
which is the average number of {\em ordered} pairs of codewords with
Euclidean distance $\delta_d$.
\begin{definition}
The {\em Euclidean distance enumerating function} of a general code
$\mathcal{C}(n,M)$ is defined as
\begin{equation}\label{Ad}
A(X) \stackrel{\Delta}{=} \sum_{\delta_d}A_{\delta_d}X^{\delta_d^2},
\end{equation}
where $X$ is a dummy variable and the summation is over all possible
distance $\delta_d$. For a general code, there exist at most
$\binom{M}{2}$ non-zero coefficients $\{A_{\delta_d}\}$, which is
referred to as the Euclidean distance spectrum.
\end{definition}

To derive tangential bounds, we also need another
distance spectrum for general codes. Given a codeword
$\underline{s}$ with energy $\delta_{d_1}^2$, we denote
$B_{\delta_{d_1},\delta_{d_2},\delta_d|\underline{s}}$ the number of
codewords $\underline{\hat{s}}$ having energy $\delta_{d_2}^2$ and
the Euclidean distance $\delta_d$ with $\underline{s}$. We define
\begin{equation}\label{Bd_Bds}
B_{\delta_{d_1},\delta_{d_2},\delta_d} \stackrel{\Delta}{=}
\sum_{\underline{s}}{\rm
Pr}\{\underline{s}\}B_{\delta_{d_1},\delta_{d_2},\delta_d|\underline{s}},
\end{equation}
which is the average number of {\em ordered} pairs of codewords with
the Euclidean distance $\delta_d$ and energies $\delta_{d_1}^2$ and
$\delta_{d_2}^2$, respectively.

\begin{definition}
The {\em triangle Euclidean distance enumerating function} of a
general code $\mathcal{C}(n,M)$ is defined as
\begin{equation}\label{GF}
B(X,Y,Z) \stackrel{\Delta}{=}
\sum_{\delta_{d_1},\delta_{d_2},\delta_d}B_{\delta_{d_1},\delta_{d_2},\delta_d}X^{\delta_{d_1}^2}Y^{\delta_{d_2}^2}Z^{\delta_d^2},
\end{equation}
where $X,Y,Z$ are three dummy variables. We call
$\{B_{\delta_{d_1},\delta_{d_2},\delta_d}\}$ the triangle Euclidean
distance spectrum of the given code.
\end{definition}

\subsection{The Conventional Union Bound}
Suppose that a codeword $\underline{s}=(s_0, s_1, \cdots, s_{n-1})
\in \mathcal{C}(n,M)$ is transmitted over an AWGN channel. Let
$\underline{y} = {\underline s} + {\underline z}$ be the received
vector, where $\underline z$ is a vector of independent Gaussian
random variables with zero mean and variance $\sigma^2$. For AWGN
channels, the maximum-likelihood~(ML) decoding is equivalent to
finding the nearest codeword $\hat{\underline s}$ to $\underline y$.
The decoding error probability ${\rm Pr} \{E\}$ is
\begin{equation}\label{Error_Pro}
{\rm Pr} \{E\}=\sum_{\underline{s}}{\rm Pr}\{\underline{s}\}{\rm
Pr}\{E|\underline{s}\},
\end{equation}
where ${\rm Pr}\{E|\underline{s}\}$ is the conditional decoding
error probability when transmitting $\underline{s}$ over the
channel. As usual, we assume that each codeword $\underline{s}$ is
transmitted with equal probability, that is ${\rm
Pr}\{\underline{s}\}=1/M$. With this assumption, the code rate is
$\frac{\log M}{n}$ and the signal-to-noise ratio (SNR) is
$\frac{\sum_{\underline{s}}\|\underline{s}\|^2}{nM\sigma^2}$.


The conventional union bound on the ML decoding error probability of
a general code $\mathcal{C}(n,M)$ is
\begin{eqnarray}\label{UB}
{\rm Pr} \{E\}&=&\sum_{\underline{s}}{\rm Pr}\{\underline{s}\}{\rm
Pr}\{E|\underline{s}\}\nonumber\\
&\leq& \sum_{\underline{s}}{\rm Pr}\{\underline{s}\}\sum_{\delta_d}A_{\delta_d|\underline{s}}Q\left(\frac{\delta_d}{2\sigma}\right)\nonumber\\
&=& \sum_{\delta_d}\sum_{\underline{s}}{\rm
Pr}\{\underline{s}\}A_{\delta_d|\underline{s}}Q\left(\frac{\delta_d}{2\sigma}\right)\nonumber\\
&=&
\sum_{\delta_d}A_{\delta_d}Q\left(\frac{\delta_d}{2\sigma}\right),
\end{eqnarray}
where $Q\left(\frac{\delta_d}{2\sigma}\right)$ is the pair-wise
error probability with
\begin{equation}\label{Qfunc}
Q(x)\stackrel{\Delta}{=}\int
_{x}^{+\infty}\frac{1}{\sqrt{2\pi}}e^{-\frac{z^{2}}{2}}\,{\rm d}z.
\end{equation}
The union bound is simple since it involves only the $Q$-function
and does not require the code structure other than the Euclidean
distance spectrum. However, the union bound is loose and even
diverges in the low-SNR region. One way to solve this issue is to
use the GFBT
\begin{equation}\label{GFBT}
  {\rm Pr} \{E|\underline{s}\} \leq {\rm Pr} \{E,\underline{y}\in\mathcal{R}|\underline{s}\} + {\rm Pr}\{\underline{y}\notin \mathcal{R}|\underline{s}\},
\end{equation}
where $E$ denotes the conditional error event, $\underline{y}$
denotes the received signal vector, and $\mathcal{R}$ denotes an
arbitrary region around the transmitted signal vector
$\underline{s}$. The first term in the right hand side~(RHS)
of~(\ref{GFBT}) is usually bounded by the conditional union bound,
while the second term in the RHS of ~(\ref{GFBT}) represents the
probability of the event that the received vector $\underline{y}$
falls outside the region $\mathcal{R}$, which is considered to be
decoded incorrectly even if it may not fall outside the Voronoi
region~\cite{Agrell96}~\cite{Agrell98} of the transmitted codeword.
For convenience, we call~(\ref{GFBT}) {\em $\mathcal{R}$-bound}.
Intuitively, the more similar the region $\mathcal{R}$ is to the
Voronoi region of the transmitted signal vector, the tighter the
$\mathcal{R}$-bound is. Therefore, both the shape and the size of
the region $\mathcal{R}$ are critical to GFBT. Given the region's
shape, one can optimize its size to obtain the tightest
$\mathcal{R}$-bound. Different from most existing works, where the
size of $\mathcal{R}$ is optimized by setting to be zero the partial
derivative of the bound with respect to a parameter~(specifying the
size), we will propose an alternative method by introducing nested
Gallager's regions in the subsection~\ref{linear_code}.

\subsection{Binary Linear Codes}

For a binary linear block code $\mathcal{C}(n,M)$ of dimension
$k=\log_2M$, length $n$, and minimum Hamming distance $d_{\min}$,
suppose that a codeword $\underline{c}$ is modulated by binary phase
shift keying (BPSK), resulting in a bipolar signal vector
$\underline{s}$ with $s_t=1-2c_t$ for $0\leq t\leq n-1$. Without
loss of generality, we assume that the code $\mathcal{C}$ has at
least three non-zero codewords, i.e., its dimension $k>1$, and the
transmitted codeword is the all-zero codeword $\underline{c}^{(0)}$
(with bipolar image ${\underline{s}^{(0)}}$). Let
${\underline{\hat{c}}}$~(with bipolar image ${\underline{\hat{s}}}$)
be a codeword of Hamming weight $d$, then the Euclidean distance
between $\underline{s}^{(0)}$ and $\underline{\hat{s}}$ is
$\delta_d=2\sqrt{d}$. We define
\begin{equation}\label{Ad2}
A_d \stackrel{\Delta}{=} A_{\delta_d|\underline{s}^{(0)}},
\end{equation}
which is the number of codewords with Hamming weight $d$. Since the
constellation of binary linear block code is geometrically uniform
and each codeword is assumed to be transmitted with equal
probability, we have
\begin{eqnarray}
 {\rm Pr} \{E\}&=& \sum_{\underline{s}} {\rm Pr}\{\underline{s}\}{\rm
 Pr}\{E|\underline{s}\}\nonumber\\
 &=& {\rm Pr}\{E|\underline{s}^{(0)}\} \nonumber\\
 &\leq& \sum_d A_d Q\left(\frac{\sqrt{d}}{\sigma}\right),
\end{eqnarray}
where $\{A_d\}$ is the weight distribution of the code
$\mathcal{C}$.

\subsection{GFBT with Parameters}\label{linear_code}
In this subsection, we will present parameterized GFBT by
introducing nested Gallager regions with parameters so that Gallager
bounds can be extended to general codes conveniently. To this end,
let $\{\mathcal{R}(r), r \in \mathcal{I} \subseteq\mathbb{R}\}$ be a
family of Gallager's regions with the same shape and parameterized
by $r\in \mathcal{I}$. For example, the nested regions can be chosen
as a family of $n$-dimensional spheres of radius $r \geq 0$ centered
at the transmitted codeword ${\underline s}$. We make the following
assumptions.

{\bf Assumptions.}
\begin{enumerate}
  \item[A1.] The regions $\{\mathcal{R}(r), r \in \mathcal{I} \subseteq\mathbb{R}\}$  are {\em nested} and their boundaries partition the whole space $\mathbb{R}^n$. That is,

       \begin{equation}\label{Assump1-1}
            \mathcal{R}(r_1) \subset \mathcal{R}(r_2)~{\rm if}~r_1  < r_2,
       \end{equation}

       \begin{equation}\label{Assump1-2}
            \partial\mathcal{R}(r_1) \bigcap \partial\mathcal{R}(r_2)=\emptyset~{\rm if}~r_1 \neq r_2,
       \end{equation}

       and

       \begin{equation}\label{Assump1-3}
            \mathbb{R}^n = \bigcup\limits_{r \in \mathcal{I}} \partial\mathcal{R}(r),
        \end{equation}
        where $\partial\mathcal{R}(r)$ denotes the boundary
        surface of the region $\mathcal{R}(r)$.

  \item[A2.] Define a functional $R: {\underline y} \mapsto r$ whenever ${\underline y} \in \partial\mathcal{R}(r)$. The randomness of the received vector $\underline y$ then induces a random variable $R$. We assume that $R$ has a probability density function~(pdf) $g(r)$.

  \item[A3.] We also assume that ${\rm Pr}\{E|{\underline y} \in \partial\mathcal{R}(r),\underline{s}\}$ can be upper-bounded by a computable upper bound $f_u(r|\underline{s})$.
\end{enumerate}

For ease of notation, we may enlarge the index set $\mathcal{I}$ to
$\mathbb{R}$ by setting $g(r) \equiv 0$ for $r\notin \mathcal{I}$.
Under the above assumptions, we have the following parameterized
GFBT~\footnote{Strictly speaking, we need one more assumption that
$f_u(r|\underline{s})$ is measurable with respect to $g(r)$.}.

\begin{proposition}\label{Proposition_PGFBT}
For any $r^*\in \mathbb{R}$,
\begin{equation}\label{PGFBT}
  {\rm Pr} \{E|\underline{s}\} \leq \int_{-\infty}^{r^*} f_u(r|\underline{s}) g(r)~{\rm d}r + \int_{r^*}^{+\infty}g(r)~{\rm d}r.
\end{equation}
\end{proposition}
\begin{IEEEproof}
\begin{eqnarray}
  {\rm Pr} \{E|\underline{s}\} &=&  {\rm Pr}\{E, \underline y \in \mathcal{R}(r^*)|\underline{s}\} + {\rm Pr}\{E, \underline y \notin \mathcal{R}(r^*)|\underline{s}\}\nonumber\\
              &\leq&  {\rm Pr}\{E, \underline y \in \mathcal{R}(r^*)|\underline{s}\} + {\rm Pr}\{\underline y \notin \mathcal{R}(r^*)|\underline{s}\}\nonumber\\
                 &\leq&  \int_{-\infty}^{r^*} f_u(r|\underline{s}) g(r)~{\rm d}r + \int_{r^*}^{+\infty}g(r)~{\rm d}r. \nonumber
\end{eqnarray}
\end{IEEEproof}

An immediate question is how to choose $r^*$ to make the above bound
as tight as possible? A natural method is to set the derivative
of~(\ref{PGFBT}) with respect to $r^*$ to be zero and then solve the
equation. In this paper, we propose an alternative method for
gaining insight into the optimal parameter.

Before presenting a necessary and sufficient condition on the
optimal parameter, we need emphasize that the computable bound
$f_u(r|\underline{s})$ may exceed one. We also assume that
$f_u(r|\underline{s})$ is non-trivial, i.e., there exists some $r$
such that $f_u(r|\underline{s}) \leq 1$. For example,
$f_u(r|\underline{s})$ can be taken as the union bound conditional
on ${\underline y} \in
\partial\mathcal{R}(r)$.

\begin{theorem}~\label{Theorem_r1}
Assume that $f_u(r|\underline{s})$ is a non-decreasing and
continuous function of $r$. Let $r_1$ be a parameter that minimizes
the upper bound as shown in~(\ref{PGFBT}). Then $r_1 =
\sup\{r\in\mathcal{I}\}$ if $f_u(r|\underline{s}) < 1$ for all $r
\in \mathcal{I}$; otherwise, $r_1$ can be taken as any solution of
$f_u(r|\underline{s}) = 1$. Furthermore, if $f_u(r|\underline{s})$
is strictly increasing in an interval $[r_{\min}, r_{\max}]$ such
that $f_u(r_{\min}|\underline{s}) < 1$ and
$f_u(r_{\max}|\underline{s}) > 1$, there exists a unique $r_1 \in
[r_{\min}, r_{\max}]$ such that $f_u(r_1|\underline{s}) = 1$.
\end{theorem}

\begin{IEEEproof}
The second part is obvious since the function $f_u(r|\underline{s})$
is strictly increasing and continuous, which is helpful for solving
numerically the equation $f_u(r|\underline{s}) = 1$.

To prove the first part, it suffices to prove that neither $r_0 <
\sup\{r\in\mathcal{I}\}$ with $f_u(r_0|\underline{s}) < 1$ nor $r_2$
with $f_u(r_2|\underline{s}) > 1$ can be optimal.

Let $r_0 < \sup\{r\in\mathcal{I}\}$ such that
$f_u(r_0|\underline{s}) < 1$. Since $f_u(r|\underline{s})$ is
continuous and $r_0 < \sup\{r\in\mathcal{I}\}$, we can find
$\mathcal{I}\ni r' > r_0$ such that $f_u(r'|\underline{s}) < 1$.
Then we have

\begin{eqnarray}
&&\lefteqn{\int_{-\infty}^{r_0} f_u(r|\underline{s}) g(r)~{\rm d}r + \int_{r_0}^{+\infty}g(r)~{\rm d}r} \nonumber\\
&=&\int_{-\infty}^{r_0} f_{u}(r|\underline{s})g(r)~{\rm d}r +
\int_{r_0}^{r'} g(r)~{\rm d}r + \int_{r'}^{+\infty}g(r)~{\rm d}r \nonumber \\
&>&\int_{-\infty}^{r_0} f_{u}(r|\underline{s}) g(r)~{\rm d}r + \int_{r_0}^{r'}f_{u}(r|\underline{s})g(r)~{\rm d}r + \int_{r'}^{+\infty}g(r)~{\rm d}r \nonumber \\
&=&\int_{-\infty}^{r'}f_{u}(r|\underline{s})g(r) ~{\rm d}r +
\int_{r'}^{+\infty}g(r)~{\rm d}r, \nonumber
\end{eqnarray}
where we have used the fact that $f_u(r|\underline{s}) < 1$ for
$r\in [r_0, r']$. This shows that $r'$ is better than $r_0$.

Suppose that $r_2$ is a parameter such that $f_u(r_2|\underline{s})
> 1$. Since $f_u(r|\underline{s})$ is continuous and non-trivial, we can find $r_1
< r_2$ such that $f_u(r_1|\underline{s}) = 1$. Then we have
\begin{eqnarray}
&&\lefteqn{\int_{-\infty}^{r_2} f_u(r|\underline{s}) g(r)~{\rm d}r + \int_{r_2}^{+\infty}g(r)~{\rm d}r} \nonumber\\
&=&\int_{-\infty}^{r_{1}} f_{u}(r|\underline{s})g(r)~{\rm d}r +
\int_{r_{1}}^{r_2} f_{u}(r|\underline{s})g(r)~{\rm d}r + \int_{r_2}^{+\infty}g(r)~{\rm d}r \nonumber \\
&>&\int_{-\infty}^{r_1} f_{u}(r|\underline{s}) g(r)~{\rm d}r + \int_{r_1}^{r_2}g(r)~{\rm d}r + \int_{r_2}^{+\infty}g(r)~{\rm d}r \nonumber \\
&=&\int_{-\infty}^{r_1}f_{u}(r|\underline{s})g(r) ~{\rm d}r +
\int_{r_1}^{+\infty}g(r)~{\rm d}r, \nonumber
\end{eqnarray}
where we have used a condition that $f_u(r|\underline{s}) > 1$ for
$r\in (r_1, r_2]$, which can be fulfilled by choosing $r_1$ to be
the maximum solution of $f_u(r|\underline{s}) = 1$. This shows that
$r_1$ is better than $r_2$.
\end{IEEEproof}

\begin{corollary}\label{corollary_r_1_SNR}
Let $f_u(r|\underline{s})$ be a non-decreasing and continuous
function of $r$. If $f_u(r|\underline{s})$ does not depend on the
SNR, then the optimal parameter $r_1$ minimizing the upper
bound~(\ref{PGFBT}) does not depend on the SNR, either.
\end{corollary}
\begin{IEEEproof}
It is an immediate result from Theorem~\ref{Theorem_r1}.
\end{IEEEproof}

Theorem~\ref{Theorem_r1} requires $f_u(r|\underline{s})$ to be a
non-decreasing and continuous function of $r$, which can be
fulfilled for several well-known bounds. Without such a condition,
we may use the following more general theorem.

\begin{theorem}~\label{Theorem_general}
For any measurable subset $\mathcal{A} \subset \mathcal{I}$, we have
\begin{equation}\label{GeneralGFBT0}
    {\rm Pr}\{E|\underline{s}\} \leq \int_{r\in \mathcal{A}} f_u(r|\underline{s})g(r)~{\rm d}r + \int_{r\notin \mathcal{A}} g(r)~{\rm d}r.
\end{equation}
Within this type, the tightest bound is
\begin{equation}\label{GeneralGFBT1}
    {\rm Pr}\{E|\underline{s}\} \leq \int_{r\in \mathcal{I}_0} f_u(r|\underline{s})g(r)~{\rm d}r + \int_{r\notin \mathcal{I}_0} g(r)~{\rm d}r,
\end{equation}
where $\mathcal{I}_0 = \{r\in \mathcal{I}| f_u(r|\underline{s}) <
1\}$. Equivalently, we have
\begin{equation}\label{GeneralGFBT2}
    {\rm Pr}\{E|\underline{s}\} \leq \int_{r\in \mathcal{I}} \min\{f_u(r|\underline{s}), 1\}g(r)~{\rm d}r.
\end{equation}
\end{theorem}

\begin{IEEEproof} Let $\mathcal{G} = \bigcup_{r\in \mathcal{A}}\partial \mathcal{R}(r)$, we have
\begin{eqnarray*}
    {\rm Pr}\{E|\underline{s}\} &\leq&  {\rm Pr}\{E, {\underline y} \in \mathcal{G}|\underline{s}\} + {\rm Pr}\{{\underline y} \notin \mathcal{G}|\underline{s}\}\\
     &=& \int_{r\in \mathcal{A}} f_u(r|\underline{s})g(r)~{\rm d}r + \int_{r\notin \mathcal{A}} g(r)~{\rm d}r.
\end{eqnarray*}
Define $\mathcal{A}_0 = \{r\in \mathcal{A}| f_u(r|\underline{s}) <
1\}$ and $\mathcal{A}_1 = \{r\in \mathcal{A}| f_u(r|\underline{s})
\geq 1\}$. Similarly, define $\mathcal{B}_0 = \{r\notin \mathcal{A}|
f_u(r|\underline{s}) < 1\}$ and $\mathcal{B}_1 = \{r\notin
\mathcal{A}| f_u(r|\underline{s}) \geq 1\}$. Noticing that
\begin{eqnarray*}
\int_{r\in \mathcal{A}} \!\!f_u(r|\underline{s})g(r)~{\rm d}r \!\!\!&\geq&\!\!\! \int_{r\in \mathcal{A}_0} \!\!f_u(r|\underline{s})g(r)~{\rm d}r  + \int_{r\in \mathcal{A}_1} \!\!g(r)~{\rm d}r\\
\int_{r\notin \mathcal{A}} \!\!g(r)~{\rm d}r \!\!\!&\geq&\!\!\!
\int_{r\in \mathcal{B}_0} \!\!f_u(r|\underline{s})g(r)~{\rm d}r  +
\int_{r\in \mathcal{B}_1} \!\!g(r)~{\rm d}r,
\end{eqnarray*}
we have
\begin{eqnarray*}
& &\int_{r\in \mathcal{A}} f_u(r|\underline{s})g(r)~{\rm d}r + \int_{r\notin \mathcal{A}} g(r)~{\rm d}r\\
&\geq& \int_{r\in \mathcal{A}_0\bigcup \mathcal{B}_0} f_u(r|\underline{s})g(r)~{\rm d}r + \int_{r\in \mathcal{A}_1\bigcup \mathcal{B}_1} g(r)~{\rm d}r\\
&=&\int_{r\in \mathcal{I}_0} f_u(r|\underline{s})g(r)~{\rm d}r + \int_{r\notin \mathcal{I}_0} g(r)~{\rm d}r\\
&=& \int_{r\in \mathcal{I}} \min\{f_u(r|\underline{s}), 1\}g(r)~{\rm
d}r.
\end{eqnarray*}

\end{IEEEproof}

\subsection{Conditional Pair-Wise Error Probabilities}
Let $\delta_d$ denote the Euclidean distance between $\underline{s}$
(the transmitted codeword) and a codeword $\underline{\hat{s}}$. The
pair-wise error probability conditional on the event $\{{\underline
y} \in
\partial \mathcal{R}(r)\}$, denoted by $p_2(r, \delta_d)$, is
\begin{eqnarray}\label{Cpairwise}
    p_2(r, \delta_d) &=& {\rm Pr}\left\{\|\underline{y}-\underline{\hat{s}}\|\leq \|\underline{y}-\underline
  s\| \mid \underline{y}\in \partial\mathcal{R}(r)\right\}\nonumber\\
  &=& \frac{\int_{\|\underline{y}-\underline{\hat{s}}\| \leq \|\underline{y}-\underline
  s\|,~~\underline{y}\in \partial\mathcal{R}(r)}f(\underline y)~{\rm d}{\underline y}}{\int_{\underline{y}\in \partial\mathcal{R}(r)}f(\underline y)~{\rm d}{\underline y}},
\end{eqnarray}
where $f({\underline y})$ is the pdf of $\underline y$. Noticing
that, different from the unconditional pair-wise error
probabilities, $p_2(r, \delta_d)$ may be zero for some $r$.

We have the following lemma.
\begin{lemma}\label{LemmaCpairwise}
Suppose that, conditional on $\underline{y}\in
\partial\mathcal{R}(r)$, the received vector $\underline y$ is
uniformly distributed over $\partial\mathcal{R}(r)$. Then the
conditional pair-wise error probability $p_2(r, \delta_d)$ does not
depend on the SNR.
\end{lemma}
\begin{IEEEproof} Since $f(\underline y)$ is constant for ${\underline y} \in \partial\mathcal{R}(r)$, we have, by canceling $f(\underline y)$ from both the numerator and the denominator of $(\ref{Cpairwise})$,
\begin{equation}\label{Cpairwise1}
    p_2(r, \delta_d) = \frac{\int_{\|\underline{y}-\underline{\hat{s}}\| \leq \|\underline{y}-\underline
  s\|, \underline{y}\in \partial\mathcal{R}(r)}~{\rm d}{\underline y}}{\int_{\underline{y}\in \partial\mathcal{R}(r)}~{\rm d}{\underline y}},
\end{equation}
which shows that the conditional pair-wise error probability can be
represented as a ratio of two ``surface area" and hence does not
depend on the SNR.
\end{IEEEproof}


\begin{theorem}\label{Theorem_r_1_SNR}
Let $f_u(r|\underline{s})$ be the conditional union bound, that is,
\begin{equation}\label{CUnionbound}
    f_u(r|\underline{s}) = \sum_{\delta_d} A_{\delta_d|\underline{s}} p_2(r, \delta_d).
\end{equation}
Suppose that, conditional on $\underline{y}\in
\partial\mathcal{R}(r)$, the received vector $\underline{y}$ is
uniformly distributed over $\partial\mathcal{R}(r)$. If
$f_u(r|\underline{s})$ is a non-decreasing and continuous function
of $r$, then the optimal parameter $r_1$ minimizing the
bound~(\ref{PGFBT}) does not depend on SNR but only on the distance
spectrum of the code.
\end{theorem}
\begin{IEEEproof}
From Lemma~\ref{LemmaCpairwise}, we know that $f_u(r|\underline{s})$
does not depend on the SNR. From Corollary~\ref{corollary_r_1_SNR},
we know that $r_1$ does not depend on the SNR.

More generally, without the condition that $f_u(r|\underline{s})$ is
a non-decreasing and continuous function of $r$, the optimal
interval $\mathcal{I}_0$ defined in Theorem~\ref{Theorem_general}
does not depend on the SNR, either.

\end{IEEEproof}

\subsection{General Framework of Parameterized GFBT}
From the above subsection, we can see that there are three main
steps to derive a parameterized GFBT.  First, choose properly nested
regions specified by a parameter. Second, find the pdf of the
parameter. Finally, find a computable upper bound on the conditional
decoding error probability given that the received vector falls on
the boundary of a parameter-specified region. The key of the third
step is to find the ``projection'' of the codewords to the boundary.
Here, the ``projection" means that the intersection between the
perpendicular bisector of the segment $\overline{\underline{s}\
\underline{\hat{s}}}$ ($\underline{s}$ and $\underline{\hat{s}}$ are
the transmitted codeword and decoded codeword, respectively) and the
boundary.

\section{Single-Parameterized Upper Bounds for General Codes}\label{sec4}

For a general code, the property of geometrical uniformity may not
hold. As a result, we can not assume a particular transmitted
codeword and must average over all conditional error probabilities.
In this section, we will first derive the conditional upper bound of
${\rm Pr} \{E|\underline{s}\}$ when transmitting the codeword
$\underline{s}$ over the channel according to the framework of the
parameterized GFBT, and then obtain the upper bound of ${\rm Pr}
\{E\}$ from~(\ref{Error_Pro}).

\begin{figure}
\centering
  \includegraphics[width=4cm]{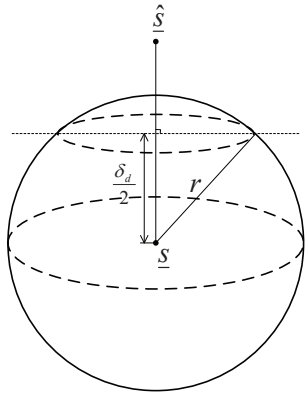}\\
  \caption{The geometric interpretation of the SB for general codes.}\label{G_SB}
  \end{figure}

\subsection{The Parameterized Sphere Bound}

\subsubsection{Nested Regions}

The parameterized SB chooses the nested regions to be a family of
$n$-dimensional spheres centered at the transmitted signal vector
$\underline{s}$, that is, $\mathcal{R}(r) = \{\underline{y}\mid
\|\underline{y}-\underline s\| \leq r\}$, where $r\geq 0$ is the
parameter. See Fig.~\ref{G_SB} for reference.

\subsubsection{Probability Density Function of the Parameter}

The pdf of the parameter is
\begin{equation}\label{SB_g}
g(r) =
\frac{2r^{n-1}e^{-\frac{r^{2}}{2\sigma^{2}}}}{2^{\frac{n}{2}}\sigma^{n}\Gamma(\frac{n}{2})},~~r\geq
0.
\end{equation}

\subsubsection{Conditional Upper Bound}
The parameterized SB chooses $f_u(r|\underline{s})$ to be the
conditional union bound when transmitting the codeword
$\underline{s}$ over the channel. Given that $||{\underline y} -
{\underline s}|| = r$, $\underline y$ is uniformly distributed over
$\partial\mathcal{R}(r)$. Hence the conditional pair-wise error
probability $p_2(r, \delta_d)$ does not depend on the SNR and can be
evaluated as the ratio of the surface area of a spherical cap to
that of the whole sphere. That is,
\begin{equation}\label{N_SB_f2}
    \!\!\!\!\!p_2(r, \delta_d)\!\! = \!\!\left\{\begin{array}{rl}
                                         \frac{\Gamma(\frac{n}{2})}{\sqrt{\pi}~\Gamma(\frac{n-1}{2})}\int_{0}^{\arccos(\frac{\delta_d}{2r})}\sin^{n-2}
\phi~{\rm d}\phi, & r > \frac{\delta_d}{2} \\
                                           0, & r \leq \frac{\delta_d}{2}
                                         \end{array}\right..
\end{equation}
Then the conditional union bound is given by
\begin{equation}\label{N_SB_f_u}
    f_u(r|\underline{s})=\sum_{\delta_d} A_{\delta_d|\underline{s}} p_2(r,
    \delta_d).
\end{equation}

\subsubsection{The Parameterized SB}
From~(\ref{GeneralGFBT2}), we have
\begin{equation}\label{con_sb}
  {\rm Pr} \{E|\underline{s}\}\leq \int_{0}^{+\infty} \min\{f_{u}(r|\underline{s}), 1\} g(r) ~{\rm
  d}r.
\end{equation}
From~(\ref{Ad_Ads}), we define
\begin{eqnarray}
f_u(r)&\triangleq&\sum_{\underline{s}}{\rm Pr}\{\underline{s}\}f_u(r|\underline{s}) \nonumber\\
&=&\sum_{\underline{s}}{\rm Pr}\{\underline{s}\}\sum_{\delta_d}
A_{\delta_d|\underline{s}} p_2(r,
    \delta_d) \nonumber\\
&=&\sum_{\delta_d}\sum_{\underline{s}}{\rm
Pr}\{\underline{s}\}A_{\delta_d|\underline{s}} p_2(r,
    \delta_d) \nonumber\\
&=&\sum_{\delta_d}A_{\delta_d} p_2(r,
    \delta_d).
\end{eqnarray}
From~(\ref{Error_Pro}), the parameterized SB for general codes can
be written as
\begin{eqnarray}\label{SB_T}
{\rm Pr}\{E\}&=&\sum_{\underline{s}}{\rm Pr}\{\underline{s}\}{\rm
Pr}\{E|\underline{s}\} \nonumber\\
&\leq& \sum_{\underline{s}}{\rm
Pr}\{\underline{s}\}\int_{0}^{+\infty}\min\left\{f_u(r|\underline{s}),1\right\}g(r)~{\rm
d}r\nonumber\\
&\leq& \int_{0}^{+\infty}\min\left\{\sum_{\underline{s}}{\rm
Pr}\{\underline{s}\}f_u(r|\underline{s}),1\right\}g(r)~{\rm d}r
\nonumber\\
&=& \int_{0}^{+\infty}\min\left\{f_u(r),1\right\}g(r)~{\rm d}r,
\end{eqnarray}
which is determined by the Euclidean distance spectrum
$\{A_{\delta_d}\}$.

\subsubsection{Reduction to Binary Linear Codes}
For binary linear codes, the transmitted codeword $\underline{s}$
can be assumed to be the all-zero codeword $\underline{s}^{(0)}$.
The Euclidean distance between a codeword $\underline{\hat{s}}$ with
Hamming weight $d$ and $\underline{s}^{(0)}$ is
$\delta_d=2\sqrt{d}$. Therefore, from~(\ref{Ad2}),~(\ref{N_SB_f2})
and~(\ref{N_SB_f_u}), the conditional union bound
$f_u(r|\underline{s}^{(0)})$ can be written as
\begin{eqnarray}
f_u(r|\underline{s}^{(0)}) &=& \sum_{\delta_d}A_{\delta_d|\underline{s}^{(0)}}p_2(r,\delta_d) \nonumber\\
&=& \sum_{1\leq d\leq n} A_d p_2(r,d),
\end{eqnarray}
where
\begin{equation}
    \!\!\!\!\!p_2(r, d)\!\! = \!\!\left\{\begin{array}{rl}
                                         \frac{\Gamma(\frac{n}{2})}{\sqrt{\pi}~\Gamma(\frac{n-1}{2})}\int_{0}^{\arccos(\frac{\sqrt{d}}{r})}\sin^{n-2}
\phi~{\rm d}\phi, & r > \sqrt{d} \\
                                           0, & r \leq \sqrt{d}
                                         \end{array}\right.,
\end{equation}
which is a non-decreasing and continuous function of $r$ such that
$p_2(0, d) = 0$ and $p_2(+\infty, d) = 1/2$. Therefore,
\begin{eqnarray} \label{SB_f_u}
    f_u(r)&=&\sum_{\underline{s}}{\rm Pr}\{\underline{s}\}f_u(r|\underline{s}) \nonumber\\
    &=& f_u(r|\underline{s}^{(0)}) \nonumber\\
    &=& \sum_{1\leq d \leq n} A_d p_2(r, d)
\end{eqnarray}
is also a non-decreasing and continuous function of $r$ such that
$f_u(0) = 0$ and $f_u(+\infty) \geq 3/2$. Furthermore, $f_u(r)$ is a
strictly increasing function in the interval $[\sqrt{d_{\min}},
+\infty)$ with $f_u(\sqrt{d_{\min}}) = 0$. Hence there exists a
unique $r_1$ satisfying

\begin{equation}\label{SB_opt}
    \sum_{1\leq d \leq n} A_d p_2(r, d) = 1,
\end{equation}
which is equivalent to that given in~\cite[(3.48)]{Sason06} by
noticing that $p_2(r, d) = 0$ for $d > r^2$.

The parameterized SB for binary linear codes can be written as
\begin{eqnarray}\label{SB}
  {\rm Pr} \{E\}&\leq& \int_{0}^{r_{1}}f_{u}(r) g(r)  ~{\rm d}r +
\int_{r_{1}}^{+\infty}g(r)~{\rm d}r\nonumber\\
&=& \int_{0}^{+\infty} \min\{f_{u}(r), 1\} g(r) ~{\rm d}r,
\end{eqnarray}
where $g(r)$ and $f_u(r)$ are given in~(\ref{SB_g})
and~(\ref{SB_f_u}), respectively. The optimal parameter $r_{1}$ is
given by solving the equation~(\ref{SB_opt}), which does not depend
on the SNR. It can be seen that~(\ref{SB}) is exactly the sphere
bound of Kasami {\em et al}~\cite{Kasami92}\cite{Kasami93}. It can
also be proved that~(\ref{SB}) is equivalent to that given
in~\cite[(3.45)-(3.48)]{Sason06}. Firstly, we have shown that the
optimal radius $r_1$ satisfies~(\ref{SB_opt}), which is equivalent
to that given in~\cite[(3.48)]{Sason06}. Secondly, by changing
variables, $z_{1} = r\cos\phi$ and $y = r^{2}$, it can be verified
that~(\ref{SB}) is equivalent to that given
in~\cite[Sec.3.2.5]{Sason06}.

\begin{figure}
\centering
  \includegraphics[width=6cm]{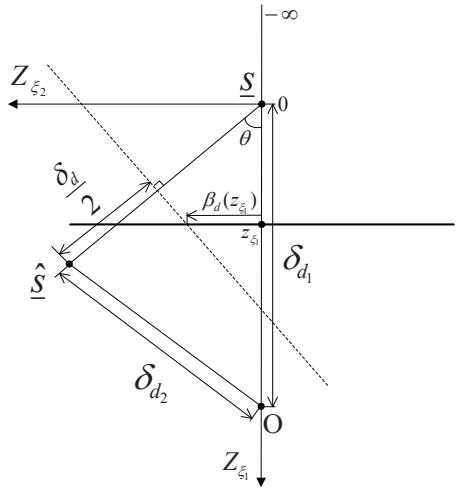}\\
  \caption{The geometric interpretation of the TB and TSB for general codes.}\label{G_TSB}
\end{figure}

\subsection{The Parameterized Tangential Bound}
In the derivation of the TB and TSB for binary codes, the
equal-energy property plays a critical role. In the rest of this
section, we show that the framework of the parameterized GFBT helps
us to generalize the TB and TSB to general codes without the
equal-energy property.

The AWGN sample $\underline z$ can be separated by projection as a
radial component $z_{\xi_1}$ and $n-1$ tangential~(orthogonal)
components $\{z_{\xi_i}, 2\leq i \leq n\}$. Specifically, we set
$z_{\xi_1}$ to be the inner product of $\underline z$ and
$-{\underline s}/\delta_{d_1}$, where $\delta_{d_1}^2$ is the energy
of $\underline s$. When considering the pair-wise error probability,
we assume that $z_{\xi_2}$ is the component that lies in the plane
determined by ${\underline s}$ and $\underline{\hat{s}}$. See
Fig.~\ref{G_TSB} for reference.

\subsubsection{Nested Regions}

The parameterized TB chooses the nested regions to be a family of
half-spaces $Z_{\xi_1} \leq z_{\xi_1}$, where $z_{\xi_1} \in
\mathbb{R}$ is the parameter. See Fig.~\ref{G_TSB} for reference.

\subsubsection{Probability Density Function of the Parameter}

The pdf of the parameter is
\begin{equation}\label{TB_g}
    g(z_{\xi_1}) = \frac{1}{\sqrt{2\pi}\sigma}e^{-\frac{z_{\xi_1}^2}{2\sigma^2}}.
\end{equation}

\subsubsection{Conditional Upper Bound}
The parameterized TB chooses $f_u(z_{\xi_1}|\underline{s})$ to be
the conditional union bound when transmitting the codeword
$\underline{s}$ over the channel. Given that $Z_{\xi_1} =
z_{\xi_1}$, the conditional pair-wise error probability is given by
\begin{equation} \label{N_TB_f2}
p_2(z_{\xi_1},\delta_{d_1},\delta_{d_2},\delta_d)=\int_{\beta_{d}(z_{\xi_1})}^{+\infty}\frac{1}{\sqrt{2\pi}\sigma}e^{-\frac{z_{\xi_2}^2}{2\sigma^2}}~{\rm
d}z_{\xi_2},
\end{equation}
where
\begin{equation}
\beta_{d}(z_{\xi_1})=\frac{\delta_d-2z_{\xi_1}\cos{\theta}}{2\sin{\theta}},
\end{equation}
and
\begin{equation}
\theta
=\arccos\left(\frac{\delta_{d_1}^2+\delta_{d}^2-\delta_{d_2}^2}{2\delta_{d_1}\delta_d}\right).
\end{equation}
Then the conditional union bound is given by
\begin{equation} \label{N_TB_f_u}
  f_u(z_{\xi_1}|\underline{s})=\sum_{\delta_{d_1},\delta_{d_2},\delta_d}B_{\delta_{d_1},\delta_{d_2},\delta_d|\underline{s}} p_2(z_{\xi_1},\delta_{d_1},\delta_{d_2},
  \delta_d).
\end{equation}

\subsubsection{The Parameterized TB}
From~(\ref{GeneralGFBT2}), we have
\begin{equation}
  {\rm Pr} \{E|\underline{s}\}\leq \int_{-\infty}^{+\infty} \min\{f_{u}(z_{\xi_1}|\underline{s}), 1\} g(z_{\xi_1}) ~{\rm
  d}z_{\xi_1}.
\end{equation}
From~(\ref{Bd_Bds}), we define
\begin{eqnarray}
f_u(z_{\xi_1})&\triangleq&\sum_{\underline{s}}{\rm
Pr}\{\underline{s}\}f_u(z_{\xi_1}|\underline{s}) \nonumber\\
&=&\sum_{\underline{s}}{\rm
Pr}\{\underline{s}\}\sum_{\delta_{d_1},\delta_{d_2},\delta_d}
B_{\delta_{d_1},\delta_{d_2},\delta_d|\underline{s}} p_2(z_{\xi_1},\delta_{d_1},\delta_{d_2},\delta_d) \nonumber\\
&=&\sum_{\delta_{d_1},\delta_{d_2},\delta_d}B_{\delta_{d_1},\delta_{d_2},\delta_d}
p_2(z_{\xi_1},\delta_{d_1},\delta_{d_2},\delta_d).
\end{eqnarray}
From~(\ref{Error_Pro}), the parameterized TB for general codes can
be written as
\begin{eqnarray}\label{TB_T}
{\rm Pr}\{E\}&=&\sum_{\underline{s}}{\rm Pr}\{\underline{s}\}{\rm
Pr}\{E|\underline{s}\} \nonumber\\
&\leq& \sum_{\underline{s}}{\rm Pr}\{\underline{s}\}
\int_{-\infty}^{+\infty} \min\{f_u(z_{\xi_1}|\underline{s}), 1\}
g(z_{\xi_1})~{\rm d}z_{\xi_1}\nonumber\\
&\leq& \int_{-\infty}^{+\infty} \min\left\{\sum_{\underline{s}}{\rm
Pr}\{\underline{s}\}f_u(z_{\xi_1}|\underline{s}), 1\right\}
g(z_{\xi_1})~{\rm d}z_{\xi_1}\nonumber\\
&=& \int_{-\infty}^{+\infty} \min\left\{f_u(z_{\xi_1}), 1\right\}
g(z_{\xi_1})~{\rm d}z_{\xi_1},
\end{eqnarray}
which is determined by the triangle Euclidean distance spectrum
$\{B_{\delta_{d_1},\delta_{d_2},\delta_d}\}$.

\begin{figure}
\centering
  \includegraphics[width=6cm]{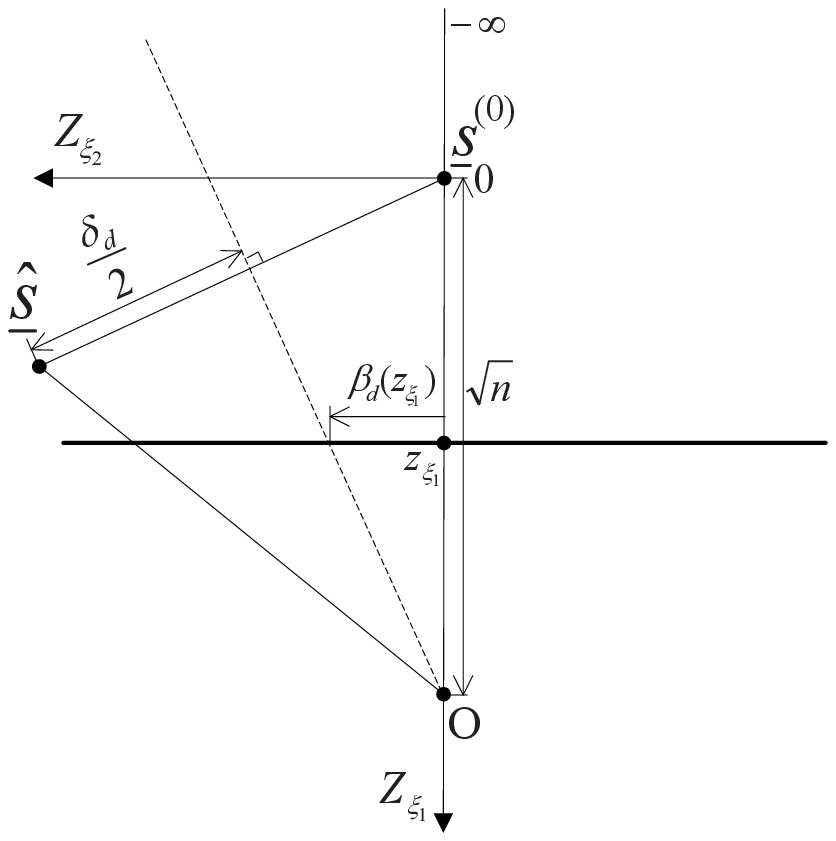}\\
  \caption{The geometric interpretation of the TB and TSB for binary linear codes.}\label{B_TSB}
\end{figure}

\subsubsection{Reduction to Binary Linear Codes}
Similarly, for binary linear codes, the transmitted codeword
$\underline{s}$ can be assumed to be the all-zero codeword
$\underline{s}^{(0)}$. The Euclidean distance between a codeword
$\underline{\hat{s}}$ with Hamming weight $d$ and energy
$\delta_{d_2}^2$ and $\underline{s}^{(0)}$ with energy
$\delta_{d_1}^2$ is $\delta_d=2\sqrt{d}$. Note that
$\delta_{d_1}=\delta_{d_2}=\sqrt{n}$, so
$B_{\delta_{d_1},\delta_{d_2},\delta_d|\underline{s}^{(0)}}=A_{\delta_d|\underline{s}^{(0)}}$.
See Fig.~\ref{B_TSB} for reference. Therefore,
from~(\ref{Ad2}),~(\ref{N_TB_f2}) and~(\ref{N_TB_f_u}), the
conditional union bound $f_u(z_{\xi_1}|\underline{s}^{(0)})$ can be
written as
\begin{eqnarray}
f_u(z_{\xi_1}|\underline{s}^{(0)}) &=&
\sum_{\delta_{d_1},\delta_{d_2},\delta_d}
B_{\delta_{d_1},\delta_{d_2},\delta_d|\underline{s}^{(0)}}
p_2(z_{\xi_1},\delta_{d_1},\delta_{d_2},\delta_d)\nonumber\\
&=& \sum_{\delta_d} A_{\delta_d|\underline{s}^{(0)}}
p_2(z_{\xi_1},\sqrt{n},\sqrt{n},\delta_d) \nonumber\\
&=& \sum_{1\leq d \leq n} A_d p_2(z_{\xi_1},d),
\end{eqnarray}
where
\begin{equation}
    p_2(z_{\xi_1}, d) =\int_{\beta_d(z_{\xi_1})}^{+\infty}\frac{1}{\sqrt{2\pi}\sigma}e^{-\frac{z_{\xi_2}^2}{2\sigma^2}}~{\rm
    d}z_{\xi_2},
\end{equation}
and
\begin{equation}
\beta_d(z_{\xi_1})=\frac{\sqrt{d}(\sqrt{n}-z_{\xi_1})}{\sqrt{n-d}}.
\end{equation}
$p_2(z_{\xi_1}, d)$ is a strictly increasing and continuous function
of $z_{\xi_1}$ such that $p_2(-\infty, d) = 0$ and $p_2(\sqrt{n}, d)
= 1/2$. Therefore,
\begin{eqnarray} \label{TB_f_u}
    f_u(z_{\xi_1})&=&\sum_{\underline{s}}{\rm Pr}\{\underline{s}\}f_u(z_{\xi_1}|\underline{s}) \nonumber\\
    &=& f_u(z_{\xi_1}|\underline{s}^{(0)}) \nonumber\\
    &=& \sum_{1\leq d \leq n} A_d p_2(z_{\xi_1}, d)
\end{eqnarray}
is also a strictly increasing and continuous function of $z_{\xi_1}$
such that $f_u(-\infty) = 0$ and $f_u(\sqrt{n}) \geq 3/2$. Hence
there exists a unique solution $z_{\xi_1}^* \leq \sqrt{n}$
satisfying

\begin{equation}\label{TB_opt}
  \sum_{d=1}^{n}A_{d} p_2(z_{\xi_1}, d) = 1,
 \end{equation}
which is equivalent to that given in~\cite[(3.22)]{Sason06} by
noticing that $p_2(z_{\xi_1}, d) =
Q\left(\frac{\sqrt{d}(\sqrt{n}-z_{\xi_1})}{\sigma\sqrt{n-d}}\right)$
and $d = \delta_d^2 / 4$.

The parameterized TB for binary linear codes can be written as
\begin{eqnarray}\label{TB}
{\rm Pr} \{E\} &\leq& \int_{-\infty}^{z_{\xi_1}^*} \!\!\!
f_{u}(z_{\xi_1})  g(z_{\xi_1}) ~{\rm d}z_{\xi_1} +
\int_{z_{\xi_1}^*}^{+\infty}
\!\!\!\!\!g(z_{\xi_1})~{\rm d}z_{\xi_1}\nonumber\\
&=&\int_{-\infty}^{+\infty} \min\{f_u(z_{\xi_1}), 1\}
g(z_{\xi_1})~{\rm d}z_{\xi_1},
\end{eqnarray}
where $g(z_{\xi_1})$ and $f_u(z_{\xi_1})$ are given in~(\ref{TB_g})
and~(\ref{TB_f_u}), respectively. The optimal parameter
$z_{\xi_1}^*$ is given by solving the equation~(\ref{TB_opt}). It
can be shown that~(\ref{TB}) is equivalent to that given
in~\cite[(3.21)]{Sason06}.

\subsection{The Parameterized Tangential-Sphere Bound}
Assume that $n\geq 3$.
\subsubsection{Nested Regions}
Again, the parameterized TSB chooses the nested regions to be a
family of half-spaces $Z_{\xi_1} \leq z_{\xi_1}$, where $z_{\xi_1}
\in \mathbb{R}$ is the parameter.

\subsubsection{Probability Density Function of the Parameter}

The pdf of the parameter is
\begin{equation}\label{TSB_g}
    g(z_{\xi_1}) = \frac{1}{\sqrt{2\pi}\sigma}e^{-\frac{z_{\xi_1}^2}{2\sigma^2}}.
\end{equation}

\subsubsection{Conditional Upper Bound}
Different from the parameterized TB, the parameterized TSB chooses
$f_u(z_{\xi_1}|\underline{s})$ to be the conditional sphere bound
when transmitting the codeword $\underline{s}$ over the channel. The
conditional sphere bound given that $Z_{\xi_1} = z_{\xi_1}$ can be
derived as follows.

Let $\mathcal{R}(r)$ be the $(n-1)$-dimensional sphere of radius
$r>0$ which is centered at $(1-z_{\xi_1}/\delta_{d_1})\underline{s}$
and located inside the hyper-plane $Z_{\xi_1} = z_{\xi_1}$. See
Fig.~\ref{G_TSB} for reference.

Given that the received vector $\underline{y}$ falls on the
$(n-1)$-dimensional sphere $\partial \mathcal{R}(r)$ in the
hyper-plane $Z_{\xi_1} = z_{\xi_1}$, the conditional pair-wise error
probability is
\begin{equation}\label{N_TSB_f2}
p_2(z_{\xi_1}, r,\delta_{d_1},\delta_{d_2}, \delta_d) =
\left\{\begin{array}{rl}
                                         \frac{\Gamma(\frac{n-1}{2})}{\sqrt{\pi}~\Gamma(\frac{n-2}{2})}\int_{0}^{\arccos(\frac{\beta_{d}(z_{\xi_1})}{r})}\sin^{n-3}
    \phi~{\rm d}\phi, & r \geq \beta_{d}(z_{\xi_1}), \beta_{d}(z_{\xi_1})> 0 \\
    0, & r < \beta_{d}(z_{\xi_1}), \beta_{d}(z_{\xi_1})> 0 \\
                                         1-\frac{\Gamma(\frac{n-1}{2})}{\sqrt{\pi}~\Gamma(\frac{n-2}{2})}\int_{0}^{\arccos(\frac{|\beta_{d}(z_{\xi_1})|}{r})}\sin^{n-3}
    \phi~{\rm d}\phi, & r \geq |\beta_{d}(z_{\xi_1})|, \beta_{d}(z_{\xi_1})\leq 0 \\
                                             1,
    & r < |\beta_{d}(z_{\xi_1})|, \beta_{d}(z_{\xi_1})\leq 0
                                         \end{array}\right.,
\end{equation}
where
\begin{equation}
\beta_{d}(z_{\xi_1})=\frac{\delta_d-2z_{\xi_1}\cos{\theta}}{2\sin{\theta}},
\end{equation}
and
\begin{equation}
\theta
=\arccos\left(\frac{\delta_{d_1}^2+\delta_{d}^2-\delta_{d_2}^2}{2\delta_{d_1}\delta_d}\right).
\end{equation}
From~(\ref{con_sb}), we have the conditional sphere bound
\begin{equation} \label{N_TSB_f_u}
f_u(z_{\xi_1}|\underline{s})=\int_{0}^{+\infty}\min\left\{f_s(z_{\xi_1},r|\underline{s}),1\right\}g_s(r)~{\rm
d}r,
\end{equation}
where
\begin{equation}
g_s(r) =
\frac{2r^{n-2}e^{-\frac{r^{2}}{2\sigma^{2}}}}{2^{\frac{n-1}{2}}\sigma^{n-1}\Gamma(\frac{n-1}{2})},~~r\geq
0,
\end{equation}
and
\begin{equation}\label{G_fs}
    f_s(z_{\xi_1},r|\underline{s}) = \sum_{\delta_{d_1},\delta_{d_2},\delta_d}
    B_{\delta_{d_1},\delta_{d_2},\delta_d|\underline{s}} p_2(z_{\xi_1},r,\delta_{d_1},\delta_{d_2},\delta_d).
\end{equation}

\subsubsection{The Parameterized TSB}
From~(\ref{GeneralGFBT2}), we have
\begin{eqnarray}
{\rm Pr} \{E|\underline{s}\} &\leq& \int_{-\infty}^{+\infty}
\min\{f_{u}(z_{\xi_1}|\underline{s}), 1\} g(z_{\xi_1}) ~{\rm
  d}z_{\xi_1}\nonumber\\
  &\leq& \int_{-\infty}^{+\infty}\min\left\{\int_{0}^{+\infty}\min\left\{f_s(z_{\xi_1},r|\underline{s}),1\right\}g_s(r)~{\rm
d}r,1\right\}g(z_{\xi_1})~{\rm d}z_{\xi_1}.\nonumber\\
\end{eqnarray}
From~(\ref{Bd_Bds}), we define
\begin{eqnarray}
f_s(z_{\xi_1},r)&\triangleq&\sum_{\underline{s}}{\rm
Pr}\{\underline{s}\}f_s(z_{\xi_1},r|\underline{s}) \nonumber\\
&=&\sum_{\underline{s}}{\rm
Pr}\{\underline{s}\}\sum_{\delta_{d_1},\delta_{d_2},\delta_d}
B_{\delta_{d_1},\delta_{d_2},\delta_d|\underline{s}}
p_2(z_{\xi_1},r,\delta_{d_1},\delta_{d_2},\delta_d) \nonumber\\
&=&\sum_{\delta_{d_1},\delta_{d_2},\delta_d}B_{\delta_{d_1},\delta_{d_2},\delta_d}
p_2(z_{\xi_1},r,\delta_{d_1},\delta_{d_2},\delta_d).
\end{eqnarray}
From~(\ref{Error_Pro}), the parameterized TSB for general codes can
be written as
\begin{eqnarray}\label{TSB_T}
{\rm Pr\{E\}} &=& \sum_{\underline{s}}{\rm Pr}\{\underline{s}\}{\rm
Pr}\{E|\underline{s}\} \nonumber\\
&\leq&\sum_{\underline{s}}{\rm
Pr}\{\underline{s}\}\int_{-\infty}^{+\infty}\min\left\{\int_{0}^{+\infty}\min\left\{f_s(z_{\xi_1},r|\underline{s}),1\right\}g_s(r)~{\rm
d}r,1\right\}g(z_{\xi_1})~{\rm
d}z_{\xi_1}\nonumber\\
&\leq&\int_{-\infty}^{+\infty}\min\left\{\int_{0}^{+\infty}\min\left\{\sum_{\underline{s}}{\rm
Pr}\left\{\underline{s}\right\}f_s(z_{\xi_1},r|\underline{s}),1\right\}g_s(r)~{\rm
d}r,1\right\}g(z_{\xi_1})~{\rm d}z_{\xi_1} \nonumber\\
&=&\int_{-\infty}^{+\infty}\min\left\{\int_{0}^{+\infty}\min\left\{f_s(z_{\xi_1},r),1\right\}g_s(r)~{\rm
d}r,1\right\}g(z_{\xi_1})~{\rm d}z_{\xi_1},
\end{eqnarray}
which is determined by the triangle Euclidean distance spectrum
$\{B_{\delta_{d_1},\delta_{d_2},\delta_d}\}$.

\subsubsection{Reduction to Binary Linear Codes}
Similarly, for binary linear codes, the transmitted codeword
$\underline{s}$ can be assumed to be the all-zero codeword
$\underline{s}^{(0)}$. The Euclidean distance between a codeword
$\underline{\hat{s}}$ with Hamming weight $d$ and energy
$\delta_{d_2}^2$ and $\underline{s}^{(0)}$ with energy
$\delta_{d_1}^2$ is $\delta_d=2\sqrt{d}$. Note that
$\delta_{d_1}=\delta_{d_2}=\sqrt{n}$, so
$B_{\delta_{d_1},\delta_{d_2},\delta_d|\underline{s}^{(0)}}=A_{\delta_d|\underline{s}^{(0)}}$.
See Fig.~\ref{B_TSB} for reference. Therefore,
from~(\ref{N_TSB_f_u}), the conditional sphere bound
$f_u(z_{\xi_1}|\underline{s}^{(0)})$ can be written as
\begin{equation}\label{TSB_f_u_s0}
f_u(z_{\xi_1}|\underline{s}^{(0)})=\int_{0}^{+\infty}\min\left\{f_s(z_{\xi_1},r|\underline{s}^{(0)}),1\right\}g_s(r)~{\rm
d}r.
\end{equation}
From~(\ref{Ad2}),~(\ref{N_TSB_f2}) and~(\ref{G_fs}), we have
\begin{eqnarray}\label{fs}
f_s(z_{\xi_1},r|\underline{s}^{(0)}) &=&
\sum_{\delta_{d_1},\delta_{d_2},\delta_d}
    B_{\delta_{d_1},\delta_{d_2},\delta_d|\underline{s}^{(0)}}
    p_2(z_{\xi_1},r,\delta_{d_1},\delta_{d_2},\delta_d) \nonumber\\
&=& \sum_{\delta_d}
    A_{\delta_d|\underline{s}^{(0)}}
    p_2(z_{\xi_1},r,\sqrt{n},\sqrt{n},\delta_d) \nonumber\\
&=& \sum_{1\leq d\leq n} A_d p_2(z_{\xi_1},r,d),
\end{eqnarray}
where
\begin{equation}\label{p2}
p_2(z_{\xi_1}, r,d) = \left\{\begin{array}{rl}
                                         \frac{\Gamma(\frac{n-1}{2})}{\sqrt{\pi}~\Gamma(\frac{n-2}{2})}\int_{0}^{\arccos(\frac{\beta_{d}(z_{\xi_1})}{r})}\sin^{n-3}
    \phi~{\rm d}\phi, & r \geq \beta_{d}(z_{\xi_1}), z_{\xi_1}< \sqrt{n} \\
    0, & r < \beta_{d}(z_{\xi_1}), z_{\xi_1}< \sqrt{n} \\
                                         1-\frac{\Gamma(\frac{n-1}{2})}{\sqrt{\pi}~\Gamma(\frac{n-2}{2})}\int_{0}^{\arccos(\frac{|\beta_{d}(z_{\xi_1})|}{r})}\sin^{n-3}
    \phi~{\rm d}\phi, & r \geq |\beta_{d}(z_{\xi_1})|, z_{\xi_1}\geq \sqrt{n} \\
                                             1,
    & r < |\beta_{d}(z_{\xi_1})|, z_{\xi_1}\geq \sqrt{n}
                                         \end{array}\right.,
\end{equation}
and
\begin{equation}\label{beta_d}
\beta_d(z_{\xi_1})=\frac{\sqrt{d}(\sqrt{n}-z_{\xi_1})}{\sqrt{n-d}}.
\end{equation}
Then
\begin{eqnarray} \label{TSB_f_u}
    f_u(z_{\xi_1})&=&\sum_{\underline{s}}{\rm Pr}\{\underline{s}\}f_u(z_{\xi_1}|\underline{s}) \nonumber\\
    &=& f_u(z_{\xi_1}|\underline{s}^{(0)}).
\end{eqnarray}

\begin{itemize}
\item[] {\em Case 1}: $Z_{\xi_1} = z_{\xi_1} \geq \sqrt{n}$. It can be shown that, given that
received vector falls on $\partial\mathcal{R}(r)$, the pair-wise
error probability is no less than 1/2. Hence the conditional union
bound is no less than 3/2. From Theorem~\ref{Theorem_r1}, we know
that the optimal radius $r_1(z_{\xi_1}) = 0$, which results in the
trivial upper bound  $f_{u}(z_{\xi_1}) \equiv 1$.
\item[] {\em Case 2}: Given that $Z_{\xi_1} = z_{\xi_1} < \sqrt{n}$, the ML decoding error probability can be
evaluated by considering an equivalent system in which each bipolar
codeword is scaled by a factor $(\sqrt{n} - z_{\xi_1})/\sqrt{n}$
before transmitted over an AWGN channel with~({\em projective})
noise $(0, Z_{\xi_2}, \cdots, Z_{\xi_n})$. The system is also
equivalent to transmission of the original codewords over an AWGN
but with scaled~({\em projective}) noise $\sqrt{n}/(\sqrt{n} -
z_{\xi_1}) (0, Z_{\xi_2}, \cdots, Z_{\xi_n})$. The latter
reformulation allows us to get the conditional sphere bound easily
since the optimal radius is independent of the SNR. From~(\ref{p2}),
given that the received signal $\underline{y}$ falls on the
$(n-1)$-dimensional sphere $\partial \mathcal{R}(r)$ in the
hyper-plane $Z_{\xi_1} = 0$, the conditional pair-wise error
probability is
        \begin{equation*}
            \!\!\!\!\!p_2(0, r, d)\!\! = \!\!\frac{\Gamma(\frac{n-1}{2})}{\sqrt{\pi}~\Gamma(\frac{n-2}{2})}\int_{0}^{\arccos(\frac{\sqrt{nd/(n-d)}}{r})}\sin^{n-3}
        \phi~{\rm d}\phi
        \end{equation*}
if $r > \sqrt{nd/(n-d)}$ and $p_2(0, r, d) = 0$ otherwise. Then we
have the conditional sphere bound

       \begin{equation}\label{TSBu}
            f_u(z_{\xi_1}) = \int_{0}^{r_{1}}  f_{s}(0,r|\underline{s}^{(0)})g_{s}(z_{\xi_1}, r) ~{\rm d}r + \int_{r_{1}}^{+\infty}g_{s}(z_{\xi_1}, r)~{\rm d}r,
       \end{equation}
       where
       \begin{equation}\label{TSBu_g}
        g_s(z_{\xi_1}, r) = \frac{2r^{n-2}e^{-\frac{r^{2}}{2\tilde{\sigma}^{2}}}}{2^{\frac{n-1}{2}}\tilde{\sigma}^{n-1}\Gamma(\frac{n-1}{2})},~~r\geq 0,
        \end{equation}
        which depends on the SNR via $\tilde \sigma = \sqrt{n}\sigma/(\sqrt{n} - z_{\xi_1})$, and
        \begin{equation}\label{TSBu_f}
            \!\!\!\!f_s(0,r|\underline{s}^{(0)}) = \!\!\!\!\!\sum_{1\leq d \leq \frac{r^2n}{r^2 + n}} A_d \frac{\Gamma(\frac{n-1}{2})}{\sqrt{\pi}~\Gamma(\frac{n-2}{2})}\int_{0}^{\arccos(\frac{\sqrt{nd/(n-d)}}{r})}\!\!\!\!\!\!\sin^{n-3}
            \phi~{\rm d}\phi,
        \end{equation}
    which is independent of $\tilde{\sigma}$, as justified previously. The optimal radius $r_1$ is the unique solution of
    \begin{equation}\label{TSBu_opt}
        \sum_{1\leq d \leq \frac{r^2n}{r^2 + n}} A_d \frac{\Gamma(\frac{n-1}{2})}{\sqrt{\pi}~\Gamma(\frac{n-2}{2})}\int_{0}^{\arccos(\frac{\sqrt{nd/(n-d)}}{r})}\!\!\!\!\!\!\sin^{n-3}
            \phi~{\rm d}\phi = 1.
    \end{equation}
Since  $r_1 < +\infty$, $f_u(z_{\xi_1}) < 1$ for all $z_{\xi_1} <
\sqrt{n}$.
\item[] {\em Summary}: We have shown that the conditional sphere upper bound satisfying that
$f_u(z_{\xi_1}) < 1 $ if $z_{\xi_1} < \sqrt{n}$ and $f_u(z_{\xi_1})
= 1 $ otherwise. Hence the optimal parameter $z_{\xi_1}^* =
\sqrt{n}$.
\end{itemize}

The parameterized TSB for binary linear codes can be written as
\begin{eqnarray}\label{TSB}
{\rm Pr} \{E\}\!\!\! &\leq&\!\!\!
\int_{-\infty}^{\sqrt{n}}\!\!\!\!\! f_{u}(z_{\xi_1})  g(z_{\xi_1})
~{\rm d}z_{\xi_1} + \int_{\sqrt{n}}^{+\infty}
\!\!\!\!\!g(z_{\xi_1})~{\rm d}z_{\xi_1},
\end{eqnarray}
where $g(z_{\xi_1})$ is given by~(\ref{TSB_g}), and $f_u(z_{\xi_1})$
is given  by~(\ref{TSBu})-(\ref{TSBu_opt}). To prove the equivalence
of~(\ref{TSB}) to the formulae given in ~\cite[Sec.3.2.1]{Sason06},
we first show that the optimal region is the same\footnote{Strictly
speaking, our derivations here show that the optimal region is a
half-cone rather than a full-cone, a fact that has never been
explicitly stated in the literatures. Once the optimal region is the
same, the two bounds must be the same except that they compute the
bounds in different ways.} as that given
in~\cite[Sec.3.2.1]{Sason06}. Noting that the optimal radius $r_1$
satisfies~(\ref{TSBu_opt}), which is equivalent to that given
in~\cite[(3.12)]{Sason06}. Back to the hyper-plane $Z_{\xi_1} =
z_{\xi_1}$, we can see that the optimal parameter is $r_1(\sqrt{n} -
z_{\xi_1})/\sqrt{n}$. This means that the optimal region is a
half-cone with the same angle as that given
in~\cite[(3.12)]{Sason06}. Then, by changing variables, $r' = r
(\sqrt{n}-z_{\xi_1}) / \sqrt{n}$, $z_{\xi_2} = r'\cos\phi$, $v =
r'^{2}-z_{\xi_2}^{2}$ and $y = r'^{2}$, it can be verified
that~(\ref{TSB}) is equivalent to that given
in~\cite[(3.10)]{Sason06}, except that the second term ${\rm
Pr}\{Z_{\xi_1} \geq \sqrt{n}\}$. This term did not appear in the
original derivation of TSB in~\cite{TSB94}, but is required as
pointed out in~\cite[Appendix A]{Sason00}.

\section{Numerical Results}\label{sec5}


As seen from Sec.~\ref{sec4}, computing the derived upper bounds
requires the Euclidian distance spectrums, which are usually
difficult to compute for general codes. In this section, we take
{\em general trellis code} as an example to compare the derived
bounds. In the case when the trellis complexity is reasonable, both
the Euclidean distance enumerating function $A(X)$ defined
in~(\ref{Ad}) and the triangle Euclidean distance enumerating
function $B(X,Y,Z)$ defined in~(\ref{GF}) are computable.

\subsection{Trellis Code}

A general code $\mathcal{C}(n,M)$ can be represented by a trellis.
The trellis can have $N$ stages. The trellis section at stage $t$
$(0\leq t\leq N-1)$, denoted by $\mathcal{B}_t$, is a subset of
$\mathcal{S}_{t} \times \mathbb{R}^{n_t} \times \mathcal{S}_{t+1}$,
where $\mathcal{S}_t$ is the state space at stage $t$ and $n_t$ is
the number of symbols associated with the $t$-th stage of the
trellis. An element $b \in \mathcal{B}_{t}$ is called a {\em branch}
and denoted by $b\stackrel{\Delta}{=} (\sigma^-(b), \ell(b),
\sigma^+(b))$, starting from a state $\sigma^-(b)\in
\mathcal{S}_{t}$, taking a label $\ell(b) \in \mathbb{R}^{n_t}$, and
ending into a state $\sigma^+(b)\in \mathcal{S}_{t+1}$. A path
through a trellis is a sequence of branches ${\underline b} = (b_0,
b_1, \cdots, b_{N-1})$ satisfying that $b_t \in \mathcal{B}_t$ and
$\sigma^-(b_{t+1}) = \sigma^+(b_{t})$.  A codeword is then
represented by a path in the sense that $\underline s = (\ell(b_0),
\ell(b_1), \cdots, \ell(b_{N-1}))$. Naturally, $\sum_{0\leq t \leq
N-1}n_t = n$, and the number of paths is $M$. Without loss of
generality, we set $\mathcal{S}_0 = \mathcal{S}_N = \{0\}$.

A trivial trellis representation of a general code
$\mathcal{C}(n,M)$ has a single starting state, a single ending
state and $M$ parallel branches, each of which is labeled by a
codeword. For most trellis algorithms, the computational complexity
is dominated by $\max |\mathcal{B}_{t}|$ and $\max
|\mathcal{S}_{t}|$, as pointed out in~\cite{McEliece96}~\cite{Ma03}.

In this paper, we assume that both $\max |\mathcal{B}_{t}|$ and
$\max |\mathcal{S}_{t}|$ are small-to-moderate. Typical examples
include terminated trellis-coded modulation
(TCM)~\cite{Ungerboeck87} and terminated intersymbol interference
(ISI) channels~\cite{Forney72}.

\subsection{Product Error Trellis}

For a general code represented by a (possibly time-invariant)
trellis, we need the product error trellis to compute the Euclidean
distance spectrums $\{A_{\delta_d}\}$ and
$\{B_{\delta_{d_1},\delta_{d_2},\delta_d}\}$. The product error
trellis has also $N$ stages. The trellis section at the $t$-th stage
is $\mathcal{B}_t\times \mathcal{B}_t$. A branch $(b_t,\hat{b}_t)\in
\mathcal{B}_t\times \mathcal{B}_t$ starts from state
$(\sigma^-(b_t),\sigma^-(\hat{b}_t))\in \mathcal{S}_t\times
\mathcal{S}_t$, takes a label $(\ell(b_t),\ell(\hat{b}_t))$, and
ends into the state $(\sigma^+(b_t),\sigma^+(\hat{b}_t))$. A pair of
codewords $(\underline{s},\underline{\hat{s}})$ correspond to a path
$((b_0,\hat{b}_0),(b_1,\hat{b}_1),\dots,(b_{N-1},\hat{b}_{N-1}))$
through the product error trellis, where $(b_0,b_1,\dots,b_{N-1})$
is the path corresponding to the codeword $\underline{s}$ and
$(\hat{b}_0,\hat{b}_1,\dots,\hat{b}_{N-1})$ is the path
corresponding to  the codeword $\underline{\hat{s}}$ . A single
error event starting at the stage $i$ and ending at the stage $j$ is
specified by a path $((b_0,\hat{b}_0),(b_1,\hat{b}_1),
\dots,(b_{N-1},\hat{b}_{N-1}))$ satisfying that

 \begin{itemize}
   \item[1.] $b_t=\hat{b}_t$ for all $t\leq i-1$,
   $\sigma^-(b_i)=\sigma^-(\hat{b}_i)$.
   \item[2.] $\sigma^+(b_t)\neq \sigma^+(\hat{b}_t)$ for all $i\leq t\leq
   j-1$, $\sigma^+(b_j)= \sigma^+(\hat{b}_j)$.
   \item[3.] $b_t=\hat{b}_t$ for all $t>j$.
 \end{itemize}

Since only single error events are required to calculate a tighter
union bound~\cite{Caire98a}\cite{Moon07}, we have the following
algorithms.

\begin{algorithm}\label{alg_A}
Compute the Euclidean distance enumerating functions.
\begin{algorithmic}[1]

\STATE Initialize $\alpha_t(p)=0,\alpha'_t(p)=0$, for $t \in
[0,N],p\in \mathcal{S}_t\times \mathcal{S}_t$. $\alpha_0((0,0))=1$.

\FOR {$t \in [0,N-1]$}
    \FOR {$b,\hat{b}\in \mathcal{B}_t$}
        \STATE $p=(\sigma^-(b),\sigma^-(\hat{b}))$
        \STATE $q=(\sigma^+(b),\sigma^+(\hat{b}))$
        \STATE $\gamma_{e}=X^{\|\ell(b)-\ell(\hat{b})\|^2}$  \label{branch2}

        \IF {$b= \hat{b}$}
            \STATE $\alpha'_{t+1}(q)\leftarrow \alpha'_{t+1}(q)+\alpha'_{t}(p)\gamma_{e}$
            \STATE $\alpha_{t+1}(q)\leftarrow \alpha_{t+1}(q)+\alpha_{t}(p)\gamma_{e}$
        \ELSE
            \IF {$\sigma^+(b)=\sigma^+(\hat{b})$}
                \STATE $\alpha'_{t+1}(q)\leftarrow \alpha'_{t+1}(q)+\alpha_{t}(p)\gamma_{e}$
            \ELSE
                \STATE $\alpha_{t+1}(q)\leftarrow \alpha_{t+1}(q)+\alpha_{t}(p)\gamma_{e}$
            \ENDIF
        \ENDIF
    \ENDFOR
\ENDFOR \STATE $A(X)=\alpha'_N((0,0))/M$
\RETURN $A(X)$        

\end{algorithmic}
{\bf Remark.} To compute the triangle Euclidean distance enumerating
function, we only need to replace $A(X)$ with $B(X, Y, Z)$ and
define
$\gamma_{e}=X^{\|\ell(b)\|^2}Y^{\|\ell(\hat{b})\|^2}Z^{\|\ell(b)-\ell(\hat{b})\|^2}$
in line~\ref{branch2}.
\end{algorithm}


\begin{figure}
\centering
  \includegraphics[width=8cm]{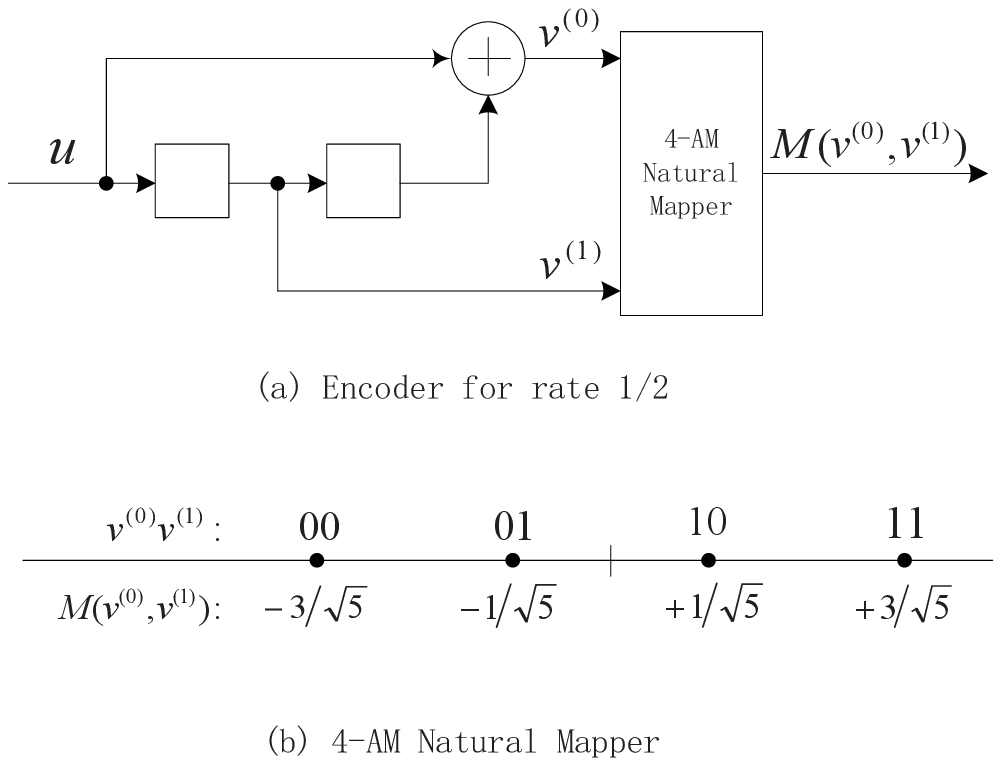}\\
  \caption{Realization of $4$-AM trellis code by means of a convolutional encoder.}\label{TCM_encoder1}
  \end{figure}

\begin{figure}
\centering
  \includegraphics[width=8cm]{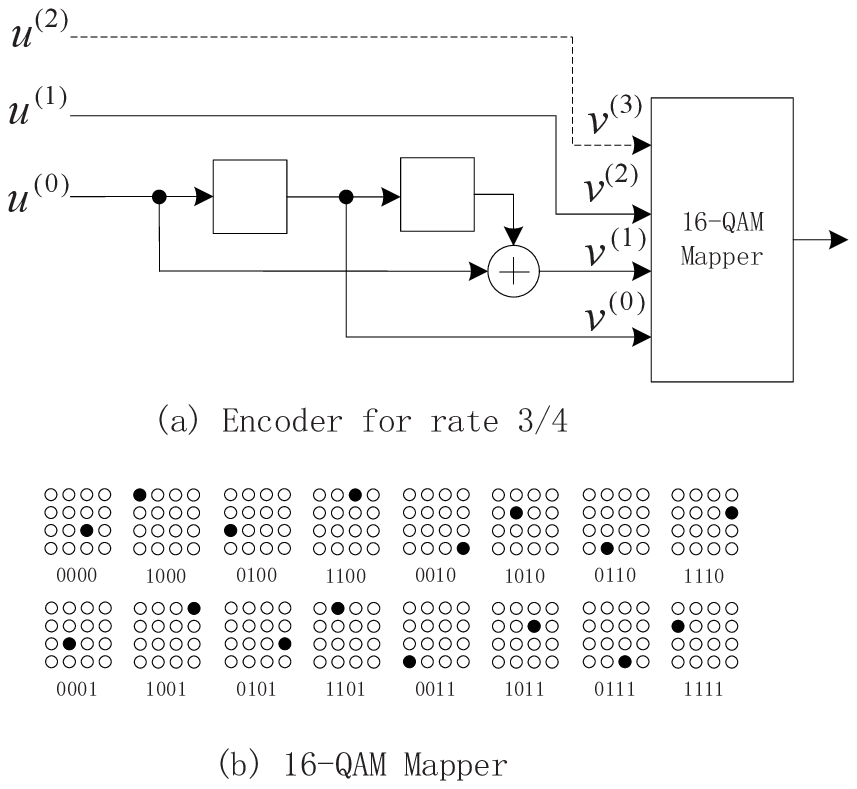}\\
  \caption{Realization of $16$-QAM trellis code by means of a minimal convolutional encoder~\cite{Ungerboeck82}.}\label{TCM_encoder2}
  \end{figure}

\begin{figure}
\centering
  \includegraphics[width=9cm]{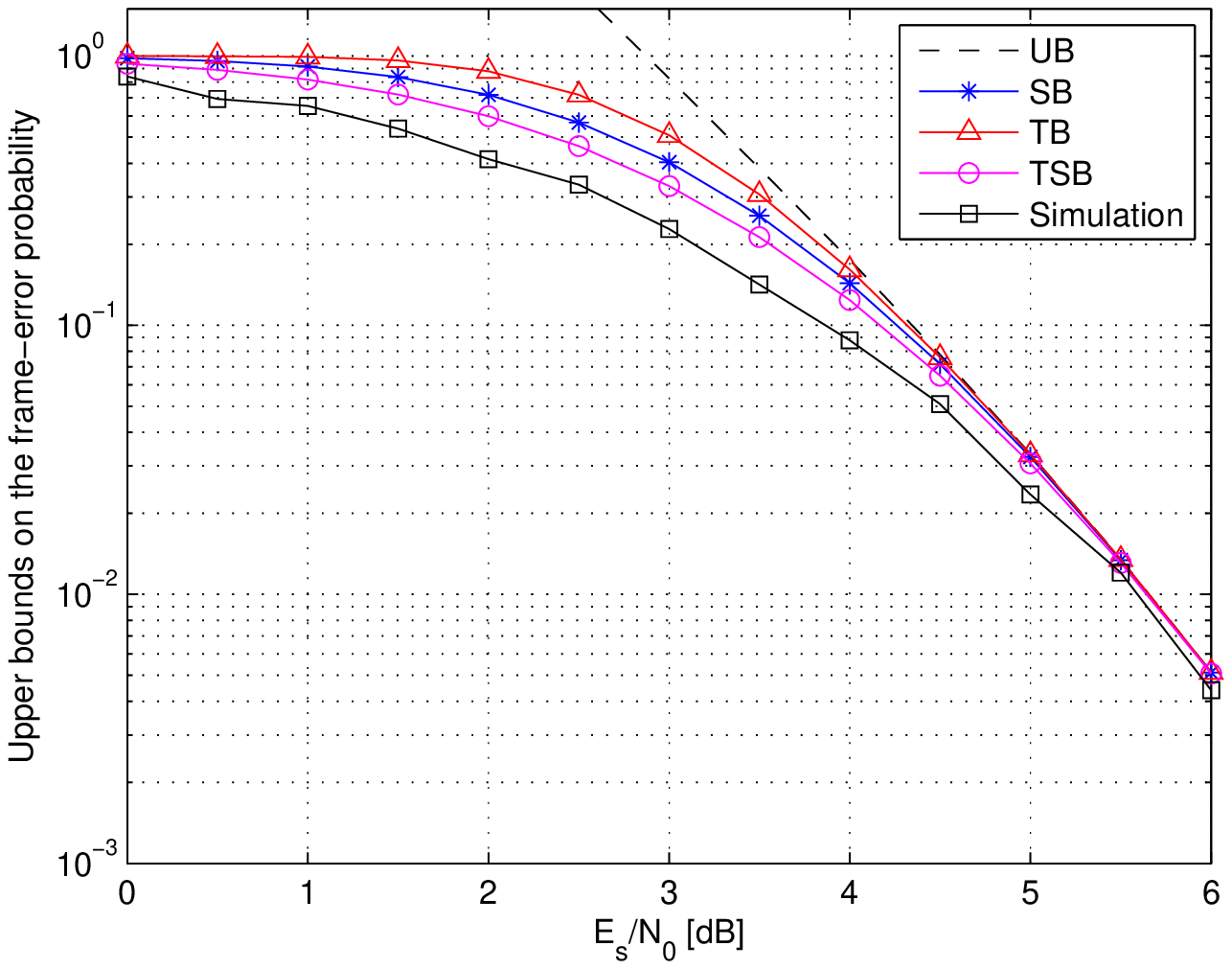}\\
  \caption{Upper bounds on the frame-error probability for the terminated trellis code $(32,2^{30})$ as shown in Fig.~\ref{TCM_encoder1}.}\label{TCM_Bound1}
  \end{figure}

\begin{figure}
\centering
  \includegraphics[width=9cm]{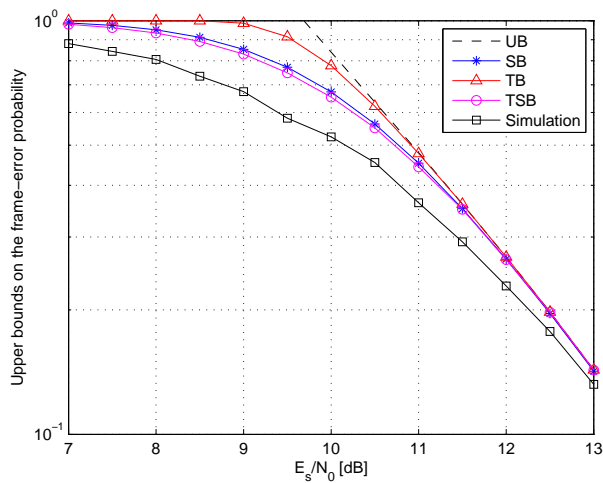}\\
  \caption{Upper bounds on the frame-error probability for the terminated trellis code $(24,2^{30})$ as shown in Fig.~\ref{TCM_encoder2}.}\label{TCM_Bound2}
  \end{figure}

\subsection{Numerical Results}


Realizations of $4$-AM and $16$-QAM trellis codes by means of
convolutional encoders are shown in Fig.~\ref{TCM_encoder1} and
Fig.~\ref{TCM_encoder2}, respectively, which result in transmitting
signals with unequal energy over AWGN channels.
From~(\ref{UB}),~(\ref{SB_T}),~(\ref{TB_T}) and~(\ref{TSB_T}), the
comparisons between the union bound, the parameterized SB, the
parameterized TB, the parameterized TSB and the simulation result on
the frame-error probability of the two terminated trellis codes are
shown in Fig.~\ref{TCM_Bound1} and Fig.~\ref{TCM_Bound2},
respectively.

\section{Conclusions}\label{conclusion}


In this paper, we have presented a general framework to investigate
Gallager's first bounding technique with a single parameter to
derive upper bounds on the ML decoding error probability of general
codes. With the proposed parameterized GFBT, the SB, the TB and the
TSB are generalized to general codes without the properties of
geometrical uniformity and equal energy. It was shown that the SB
can be calculated given that the Euclidean distance spectrum of the
code is available and that both the TB and the TSB can be calculated
given that the triangle Euclidean distance spectrum of the code is
available. When applied to binary linear codes, the triangle
distance spectrum is reduced to the conventional weight
distribution. As a result, the three generalized bounds are reduced
to the conventional ones. With the proposed parameterized GFBT, the
equation for the optimal parameter can be obtained in an intuitive
manner without resorting to the derivatives.

\small
\bibliographystyle{IEEEtran}

\begin{thebibliography}{10}
\providecommand{\url}[1]{#1} \csname url@rmstyle\endcsname
\providecommand{\newblock}{\relax} \providecommand{\bibinfo}[2]{#2}
\providecommand\BIBentrySTDinterwordspacing{\spaceskip=0pt\relax}
\providecommand\BIBentryALTinterwordstretchfactor{4}
\providecommand\BIBentryALTinterwordspacing{\spaceskip=\fontdimen2\font
plus \BIBentryALTinterwordstretchfactor\fontdimen3\font minus
  \fontdimen4\font\relax}
\providecommand\BIBforeignlanguage[2]{{%
\expandafter\ifx\csname l@#1\endcsname\relax
\typeout{** WARNING: IEEEtran.bst: No hyphenation pattern has been}%
\typeout{** loaded for the language `#1'. Using the pattern for}%
\typeout{** the default language instead.}%
\else \language=\csname l@#1\endcsname \fi #2}}

\bibitem{Sason06}
I.~Sason and S.~Shamai, ``Performance analysis of linear codes under
  maximum-likelihood decoding: {A} tutorial,'' in \emph{Foundations and Trends
  in Commun. and Inf. Theory}.\hskip 1em plus 0.5em minus 0.4em\relax Delft,
  The Netherlands: NOW, July 2006, vol.~3, no. 1-2, pp. 1--225.

\bibitem{Duman98}
T.~M. Duman and M.~Salehi, ``New performance bounds for turbo
codes,''
  \emph{IEEE Trans. Inf. Theory}, vol.~46, no.~6, pp. 717--723, June 1998.

\bibitem{Duman98a}
T.~M. Duman, ``Turbo codes and turbo coded modulation systems:
Analysis and
  performance bounds,'' Ph.D. dissertation, Elect. Comput. Eng. Dept.,
  Northeastern Univ., Boston, MA, May 1998.

\bibitem{Shulman99}
N.~Shulman and M.~Feder, ``Random coding techniques for nonrandom
codes,''
  \emph{IEEE Trans. Inf. Theory}, vol.~45, no.~6, pp. 2101--2104, September
  1999.

\bibitem{Twitto07}
M.~Twitto, I.~Sason, and S.~Shamai, ``Tightened upper bounds on the
{ML}
  decoding error probability of binary linear block codes,'' \emph{IEEE Trans.
  Inf. Theory}, vol.~53, pp. 1495--1510, April 2007.

\bibitem{Berlekamp80}
E.~R. Berlekamp, ``The technology of error correction codes,''
  \emph{Proceedings of the IEEE}, vol.~68, pp. 564--593, May 1980.

\bibitem{Kasami92}
T.~Kasami, T.~Fujiwara, T.~Takata, K.~Tomita, and S.~Lin,
``Evaluation of the
  block error probability of block modulation codes by the maximum-likelihood
  decoding for an {AWGN} channel,'' in \emph{Proc. of the 15th Symp. Inf.
  Theory and Its Applications}, Minakami, Japan, September 1992.

\bibitem{Kasami93}
T.~Kasami, T.~Fujiwara, T.~Takata, and S.~Lin, ``Evaluation of the
block error
  probability of block modulation codes by the maximum-likelihood decoding for
  an {AWGN} channel,'' in \emph{Proc. 1993 IEEE Int. Symp. Inf. Theory},
  January 1993, p.~68.

\bibitem{Sphere94}
H.~Herzberg and G.~Poltyrev, ``Techniques of bounding the
probability of
  decoding error for block coded modulation structures,'' \emph{IEEE Trans.
  Inf. Theory}, vol.~40, pp. 903--911, May 1994.

\bibitem{TSB94}
G.~Poltyrev, ``Bounds on the decoding error probability of binary
linear codes
  via their spectra,'' \emph{IEEE Trans. Inf. Theory}, vol.~40, pp. 1284--1292,
  July 1994.

\bibitem{Divsalar99}
D.~Divsalar, ``A simple tight bound on error probability of block
codes with
  application to turbo codes,'' in \emph{Proc. 1999 IEEE Commun. Theory
  Workshop}, Aptos, CA, May 1999.

\bibitem{Divsalar03}
D.~Divsalar and E.~Biglieri, ``Upper bounds to error probabilities
of coded
  systems beyond the cutoff rate,'' \emph{IEEE Trans. Commun.}, vol.~51,
  no.~12, pp. 2011--2018, December 2003.

\bibitem{Yousefi04}
S.~Yousefi and A.~K. Khandani, ``A new upper bound on the {ML}
decoding error
  probability of linear binary block codes in {AWGN} interference,'' \emph{IEEE
  Trans. Inf. Theory}, vol.~50, pp. 3026--3036, Novomber 2004.

\bibitem{Yousefi04a}
------, ``Generalized tangential sphere bound on the {ML} decoding error
  probability of linear binary block codes in {AWGN} interference,'' \emph{IEEE
  Trans. Inf. Theory}, vol.~50, pp. 2810--2815, Novomber 2004.

\bibitem{Mehrabian06}
A.~Mehrabian and S.~Yousefi, ``Improved tangential sphere bound on
the {ML}
  decoding error probability of linear binary block codes in {AWGN} and block
  fading channels,'' \emph{IEE Proc. Commun.}, vol. 153, pp. 885--893, December
  2006.

\bibitem{Ma10}
X.~Ma, C.~Li, and B.~Bai, ``Maximum likelihood decoding analysis of
{LT} codes
  over {AWGN} channels,'' in \emph{Proc. of the 6th Int. Symp. on Turbo Codes
  and Iterative Information Processing}, Brest, France, September 2010.

\bibitem{Ma11}
X.~Ma, J.~Liu, and B.~Bai, ``New techniques for upper-bounding the
{MLD}
  performance of binary linear codes,'' in \emph{Proc. 2011 IEEE Int. Symp.
  Inf. Theory}, Saint-Petersburg, Russian Federation, August 2011, pp.
  2910--2914.

\bibitem{Ma13}
------, ``New techniques for upper-bounding the {ML} decoding performance of
  binary linear codes,'' \emph{IEEE Trans.~Commun.}, vol.~61, no.~3, pp.
  842--851, Mar. 2013.

\bibitem{Caire98a}
G.~Caire and E.~Viterbo, ``Upper bound on the frame error
probability of
  terminated trellis codes,'' \emph{IEEE Commun. Lett.}, vol.~1, no.~1, pp.
  2--4, Jan. 1998.

\bibitem{Agrell96}
E.~Agrell, ``Voronoi regions for binary linear block codes,''
\emph{IEEE Trans.
  Inf. Theory}, vol.~42, pp. 310--316, January 1996.

\bibitem{Agrell98}
------, ``On the {Voronoi} neighbor ratio for binary linear block codes,''
  \emph{IEEE Trans. Inf. Theory}, vol.~44, pp. 3064--3072, Novomber 1998.

\bibitem{Sason00}
I.~Sason and S.~Shamai, ``Improved upper bounds on the {ML} decoding
error
  probability of parallel and serial concatenated turbo codes via their
  ensemble distance spectrum,'' \emph{IEEE Trans. Inf. Theory}, vol.~46, pp.
  24--47, January 2000.

\bibitem{McEliece96}
R.~J. McEliece, ``On the {BCJR} trellis for linear block codes,''
\emph{IEEE
  Trans.~Inf.~Theory}, vol.~42, pp. 1072--1092, July 1996.

\bibitem{Ma03}
X.~Ma and A.~Kav\v{c}i\'c, ``Path partition and forward-only trellis
  algorithms,'' \emph{IEEE Trans. Inf. Theory}, vol.~49, no.~1, pp. 38--52,
  Jan. 2003.

\bibitem{Ungerboeck87}
G.~Ungerboeck, ``Trellis-coded modulation with redundant signal
sets-part {I}:
  Introdution and part {II}: State of the art,'' \emph{IEEE Commun.~Mag.},
  vol.~25, pp. 5--21, Feb 1987.

\bibitem{Forney72}
G.~D. Forney~Jr., ``Maximum-likelihood sequence estimation of
digital sequences
  in the presence of intersymbol interference,'' \emph{IEEE Trans. Inf.
  Theory}, vol.~18, pp. 363--378, March 1972.

\bibitem{Moon07}
H.~Moon and D.~C. Cox, ``Improved performance upper bounds for
terminated
  convolutional codes,'' \emph{IEEE Commun. Lett.}, vol.~11, no.~6, pp.
  519--521, June 2007.

\bibitem{Ungerboeck82}
G.~Ungerboeck, ``Channel coding for multilevel/phase signals,''
\emph{IEEE
  Trans.~Inf.~Theory}, vol. IT-28, pp. 55--67, Jan. 1982.

\end{thebibliography}
\end{document}